# Extreme Metastability of Diamond and its Transformation to BC8 Post-Diamond Phase of Carbon

Kien Nguyen-Cong,[†] Jonathan T. Willman,[†] Joseph M. Gonzalez,[†] Ashley S. Williams,[†] Anatoly B. Belonoshko,[‡] Stan G. Moore,[¶] Aidan P. Thompson,[¶] Mitchell A. Wood,[¶] Jon H. Eggert,[§] Marius Millot,[§] Luis A. Zepeda-Ruiz,[§] and Ivan I. Oleynik[*,†]

[†]Department of Physics, University of South Florida, Tampa, FL 33620, United States
[‡]Department of Physics, Royal Institute of Technology, 106691 Stockholm, Sweden [¶]Sandia National Laboratories, Albuquerque, NM 87185, United States [§]Lawrence Livermore National Laboratory, Livermore, CA 94550, United States

E-mail: oleynik@usf.edu

## Abstract

Diamond possesses exceptional physical properties due to its remarkably strong carbon-carbon bonding, leading to significant resilience to structural transformations at very high pressures and temperatures. Despite several experimental attempts, syn- thesis and recovery of the theoretically predicted post-diamond BC8 phase remains elusive. Through quantum accurate, multi-million atom molecular dynamics (MD) simulations, we have uncovered the extreme metastability of diamond at very high pressures, significantly exceeding its range of thermodynamic stability. We predict the post-diamond BC8 phase to be experimentally accessible only within a narrow high pressure-temperature region of the carbon phase diagram. The diamond to BC8 trans- formation proceeds through pre-melting followed by BC8 nucleation and growth in the metastable carbon liquid. We propose a double-shock compression pathway to achieve BC8 synthesis, which is currently being explored in theory-inspired experiments at the National Ignition Facility.

**TOC_GRAPHIC**

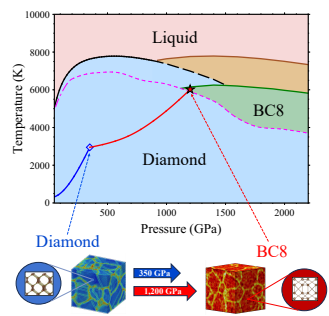

Carbon is a unique element that exhibits a variety of bonding environments at ambient conditions including single, double, and triple bonds[1] in 0-dimensional fullerenes[2], 1-D nanotubes[3], 2-D graphene/graphite[4], and 3-D diamond[5]. However, at higher pressures above 20 GPa, the energetic dominance of the strong sp$^3$ bonds reduces the structural diversity of carbon[6] to a single structure – diamond[7]. The extreme strength of carbon-carbon bonding is responsible for extraordinary properties of diamond - the highest hardness and thermal conductivity among naturally occurring materials[8], as well as exceptional chemical stability[5].

Although the properties of carbon at ambient conditions are well established[5,8], its behavior at extreme temperatures and pressures is less certain[9–13]. In the past several decades, graphite and diamond at extreme conditions were studied with the goal of uncovering the effect of pressure and temperature on structure and bonding[9–24]. Very recently, a substantial effort has been directed to investigate carbon at megabar pressures and thousands of Kelvins[25–34] to provide data for developing models of planetary interiors for carbon-rich exoplanets[27,31,35–38] as well as to optimize diamond capsules for inertial confinement fusion experiments[39–41].

Previous theoretical studies predicted a very simple high-pressure phase diagram of carbon: thermodynamically stable diamond up to 1.1 TPa, followed by a body-centered-cubic BC8 structure up to 2.3 TPa, and simple cubic and simple hexagonal structures at even higher pressures, up to 5 TPa[19,22,30]. However, recent ramp compression experiments at the National Ignition Facility (NIF), utilizing X-ray diffraction for atomic structure characterization, observed the persistence of diamond at pressures up to 2 TPa, well above the predicted diamond/BC8 phase boundary[33]. Quantum molecular dynamics simulations (QMD) using first-principles density functional theory also found no evidence of a direct diamond-to-BC8 transition due to an appreciable (~ 2eV) energy barrier separating the two crystal structures[21,24]. Therefore, these experiments and simulations raise the most important question: can BC8 be synthesized and observed in an experiment?

This work addresses the fundamental question by systematically exploring the metastability of diamond across a wide range of pressures and temperatures and predicting synthesizability domain of the BC8 post-diamond phase of carbon. Recent transformative advances in developing a machine learning Spectral Neighbor Analysis Potential (SNAP) for carbon now enable us to surpass the time- and length-scale limitations of previous QMD simulations, while achieving quantum accuracy (with less than 3 % deviation from QMD) across an extremely large pressure range (up to 5 TPa) and temperature range (up to 15,000 K)[42]. Moreover, the carbon SNAP demonstrates an excellent agreement with available experimental data including diamond shock Hugoniots[42]. By employing an efficient GPU implementation of the carbon SNAP potential on the world's fastest supercomputers we can now accurately simulate the time evolution of billions of carbon atoms under extreme PT conditions at experimental time and length scales[43].

Our first step is to construct the thermodynamic phase diagram of carbon by determining the phase boundaries: solid-liquid melting lines of diamond and BC8, calculated using the two-phase method[44], as well as the diamond/BC8 phase boundary by thermodynamic integration[45,46]. In agreement with previous experiments[26,29] and QMD simulations[30], we observe negative slopes of the diamond and BC8 melting lines at pressures above 500 and 1,400 GPa, respectively (Figure 1). This is because, above these pressures, liquid carbon becomes more dense than diamond or BC8 resulting from a significant increase in the coordination of carbon atoms (greater than four) in the liquid phase with increasing pressure[19]. Our SNAP MD simulations predict the diamond-BC8-liquid triple point to be 900 GPa and 7, 500 K, and the diamond/BC8 phase boundary of 1, 050 GPa at 0 K in agreement with previous QMD simulations[30]. Interestingly, the extension of the diamond melting line into the thermodynamic stability domain of BC8 creates a region in P −T space between the diamond and BC8 melting lines where compressed diamond could melt into a metastable supercooled carbon liquid. Since this is the P − T region of thermodynamic stability for BC8, we hypothesize a potential BC8 nucleation from a metastable carbon liquid upon compression and heating of diamond.

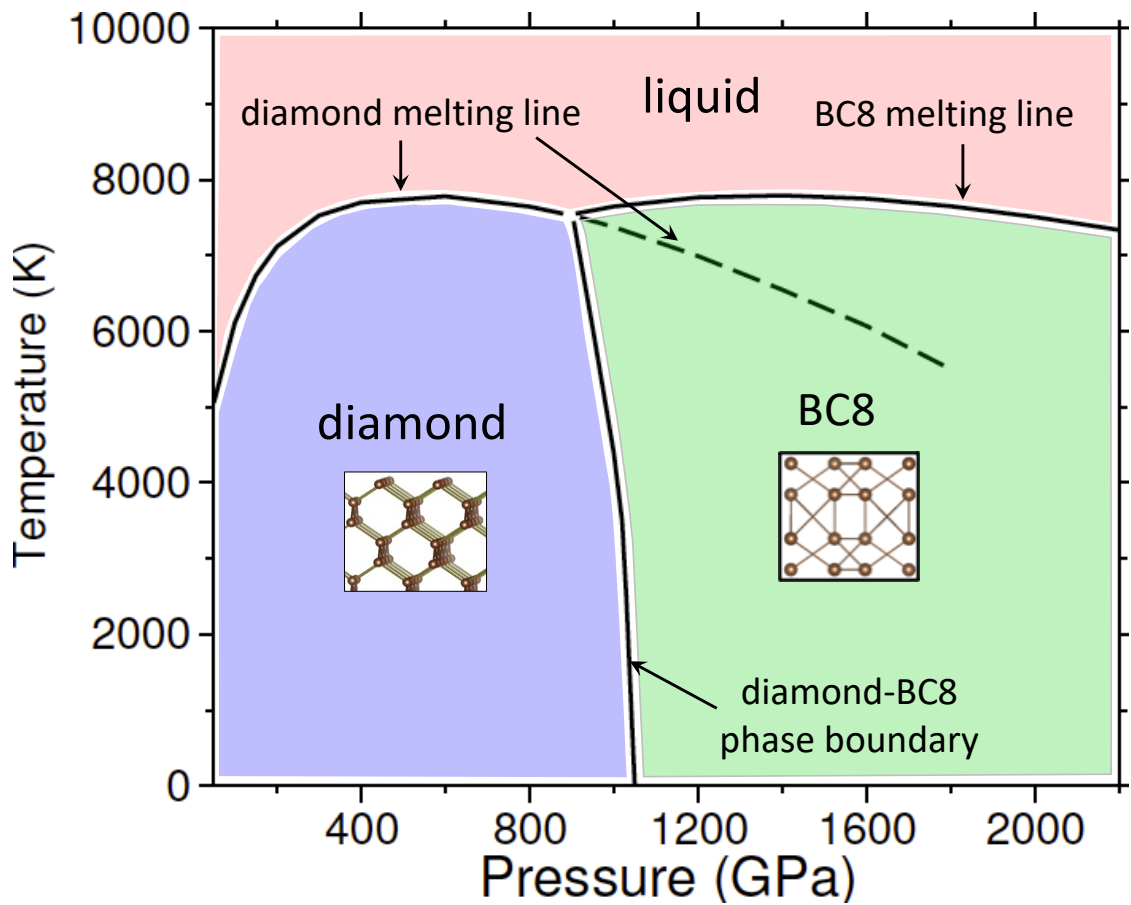

Figure 1: Phase diagram of carbon calculated in this work by quantum accurate SNAP MD. Black solid lines are the thermodynamic phase boundaries: solid-liquid diamond and BC8 melting lines, and solid-solid diamond/BC8 phase boundary. Dashed line is the extension of diamond melting line inside the range of thermodynamic stability of BC8.

To identify the range of metastability of diamond and a potential pathway to synthesize the BC8 phase, we explore the structural diversity of carbon across the entire thermodynamic P − T space of interest (Figure 1). We find that single-crystal diamond is extremely metastable within a wide range of pressures (from 0 to 2, 200 GPa) and temperatures exceeding diamond melting line (including its extension to BC8 range of thermodynamics stability) by ∼ 15−20 % (Figure S2 in the Supporting Information). Therefore, we use nano-crystalline diamond (NCD) samples rather than single crystal diamond to activate potential materials transformations seeded by grain boundaries – the defective regions between crystalline diamond grains characterized by high level of disordered bonding. The NCD samples have an average grain size ∼ 5nm (Figure S1c of the Supporting Information). In addition, NCD mimics the complex microstructure of shock-compressed single-crystal diamond, which develops inelastic deformations including local amorphization, twinning and formation of hexagonal stacking faults upon compression (Figure S3 in the Supporting Information). Importantly, NCD fuel capsules are dynamically compressed to high pressures in fusion ignition experiments at NIF[40].

Our nanosecond-long MD simulations are performed on the dense grid of (P,T) points within a wide range of pressures (up to 2, 200 GPa) and temperatures (up to 10, 000 K) utilizing million atom NCD samples. Additionally, we simulate interesting cases using billion atom NCD samples for several nanoseconds to investigate the time and length scale dependence of transformation mechanisms. Our simulation protocol at a each (P, T ) point consists of gradual cold (T = 0 K) compression of the NCD sample to the target pressure P , followed by heating to the target temperature T . Once the target (P, T ) point is reached, the behavior of the NCD sample is then simulated under isothermal-isobaric (NPT) conditions for at least 1 ns to document materials transformations at the atomic scale.

We have observed several unexpected phenomena related to NCD transformations. The resulting map of final material states, shown in Figure 2, demonstrates that NCD does indeed exhibit extreme metastability within a wide pressure range (from 0 to 2, 200 GPa) and at temperatures below 3, 500 K. As the temperature increases, appreciable local atomic mobility leads to the reconstruction of the atomic structure at grain boundaries. When temperature exceeds 5,000K, local melting at the grain boundaries is initiated across a wide range of pressures to relieve significant local strain, resulting in noticeable changes in the NCD microstructure. At even higher temperatures, approximately 1,000K below the diamond melting line, a substantial fraction of liquid forms at the grain boundaries, followed by its re-solidification into the diamond phase.

During this pre-melting process, we observe grain coarsening - the growth of larger grains at the expense of smaller ones, as depicted in the left panel of Figure 2a. Eventually, NCD undergoes a transformation into a single-crystal diamond at temperatures just below the diamond melting line (600 GPa and 7, 750 K frame in Figure 2a). The pre-melting line (indicated by the pink dash-dotted line in the center panel of Figure 2a) separates NCD states without pre-melting from higher-temperature states with a significant amount of transient metastable liquid involved in the process of grain coarsening. Similar to the diamond thermodynamic melting line, the metastable pre-melting line also extends into the $P - T$ region of thermodynamic stability of the BC8 phase.

At pressures above 1,600 GPa we observe nucleation and growth of BC8 phase in the $P - T$ region between the NCD pre-melting and melting lines (green colored region in Figure 2b). The BC8 nuclei appear within the metastable carbon liquid during the NCD pre- melting. A time progression of NCD to BC8 transformation at 2,000 GPa and 5,200K is illustrated in Figure 2b and the Movie SM-1 in the Supporting Information. Pre-melting of the NCD sample starts at the grain boundaries (10 ps frame) resulting in complete melting of the NCD sample (65 ps frame). The appearance of metastable carbon liquid is confirmed by a substantial increase in carbon diffusion within the carbon liquid fraction as shown in the left panel of Figure 2b. The first BC8 nucleus appears at ∼ 67ps immediately after completion of melting. It rapidly grows by consuming the remaining metastable liquid to form single crystal BC8, coinciding with the drop in diffusion, reduction to zero liquid faction, and almost 100 % BC8 fraction in the left panel of Figure 2b at 100 ps.

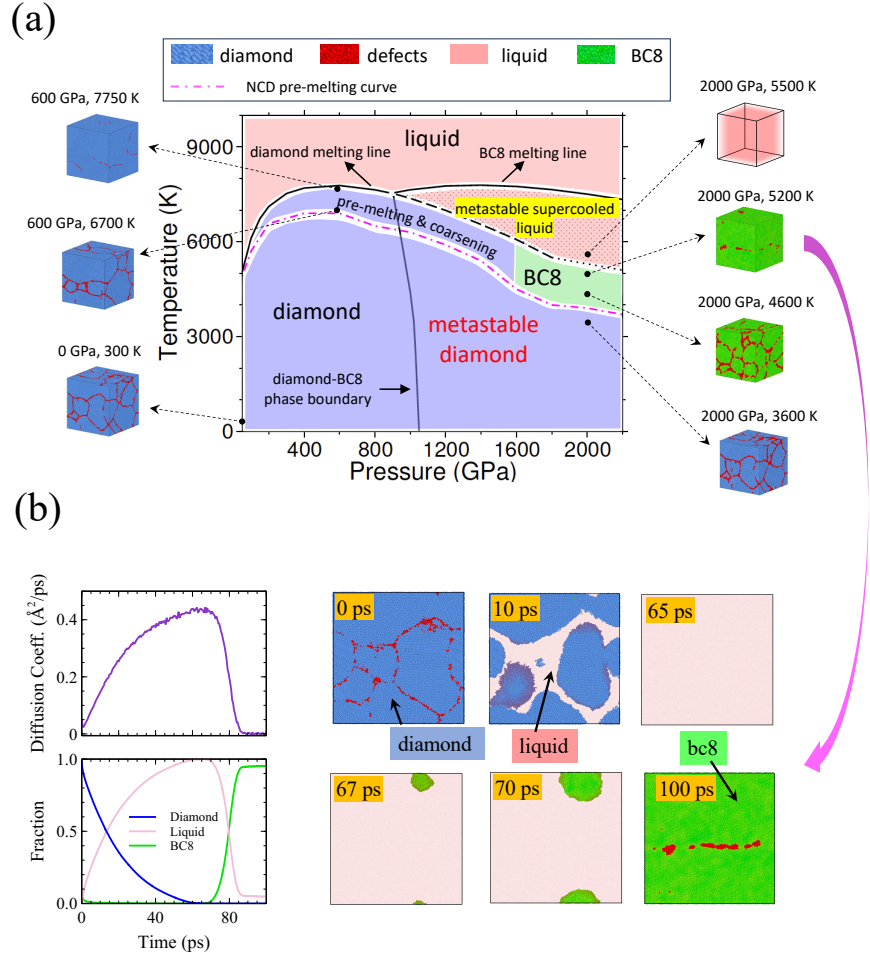

Figure 2: (a) Pressure-temperature map of the final states of NCD sample, compressed from 0 to 2,200 GPa and heated from 0 to 10,000K, obtained in a series of NPT SNAP MD simulations, each simulating the evolution of the material within 1 ns. The pre-melting line (indicated by the pink dash-dotted line) separates NCD states without pre-melting from higher-temperature states with a significant amount of transient metastable liquid involved in the process of grain coarsening. Metastable supercooled carbon liquid is observed at P, T conditions between diamond and BC8 melting lines. The frames on the left displays the initial NCD sample together with temperature-dependent final diamond states at 600 GPa. The frames on the right depict the various temperature-dependent final states at 2, 000 GPa, including NCD, BC8, and supercooled liquid. (b) Time evolution of NCD sample at specific (P, T) point, 2, 000 GPa and 5, 200 K. Left panels display diffusion coefficient and fraction of diamond, liquid (including liquid-like grain-boundary atoms) and BC8 as a function of time; right panels – transformation of initial NCD sample through full melting and solidification to BC8. See also the Movie SM-1 in the Supporting Information. Due to a significant uncertainty in identifying various atom types in a very noisy high-TP environment, some atoms are mislabeled (e.g., a very tiny fraction of BC8 atoms is seen in the left panel of Figure 2b at the beginning of the MD simulation).

We observe an appreciable temperature dependence of BC8 nucleation and growth, as well as the resultant microstructure (see frames on the right of Figure 2a). At lower temperatures, NCD does not completely melt during pre-melting, and the BC8 nuclei appear within the carbon liquid, coexisting with NCD grains. The rapid growth of BC8 grains results in the complete consumption of the carbon liquid and the formation of narrow diamond/BC8 liquid- like interfaces (Figure S4 in the Supporting Information). The NCD to BC8 interconversion is sustained by a continuous supply of liquid carbon to the thin interface on the diamond side and the fast consumption of liquid carbon at the BC8 side. As the temperature decreases from the diamond melting line towards its pre-melting line (indicated by the black dots in the Figure 2a), we observe an increase in the number of BC8 grains in the final sample. At temperatures below the diamond pre-melting line (e.g., 3,600K at 2,000 GPa), NCD samples remained intact for the entire duration of our simulations.

Although we hypothesized that the BC8 phase may nucleate within the metastable supercooled carbon liquid in the $P - T$ region between the diamond and BC8 melting lines (indicated by the black dashed and continuous lines in Figure 2a, respectively) this nucleation is not observed, at least within 1 ns of MD simulation time. To understand the reason for the lack of BC8 nucleation, we conducted another series of MD simulations to investigate the behavior of the metastable carbon liquid in the pressure interval between 1,200 and 2,200 GPa and temperatures ranging from 3,000 to 10,000K (Figure 3). To exclude any uncertainty associated with the finite cooling rate, we employ instantaneous cooling of the carbon liquid initially equilibrated at a given pressure and 10, 000 K, to bring the liquid into a supercooled state at a target temperature. Then, the material's evolution towards its final state is monitored for 1 ns. Instantaneous cooling is a physically reasonable assumption as it corresponds to an instantaneous jump in temperature in the leading shock front resulting in ultrafast melting, followed by very fast temperature drop, followed by a slow evolution behind the shock front towards the final state. See, e.g., temperature profile in the shocked diamond in Figure S7 of the Supporting Information. Using supercooled liquid as a starting state for MD simulations removes uncertainties related to the presence of partially melted diamond pieces in the metastable liquid during the BC8 nucleation when NCD is used as a starting material.

Compared to the minimum pressure of 1, 600 GPa required for the synthesis of BC8 from NCD (Figure 2a), we observe BC8 nucleation from metastable liquid carbon at the lower minimum pressure of 1,200 GPa (Figure 3a). Although the upper temperature boundary for the BC8 nucleation for both sets of simulations is approximately the same – around 5, 500 K – the lower boundary for BC8 nucleation is about 500 K higher for nucleation from NCD than from the metastable liquid across the pressure range under consideration. As in the case of direct NCD to BC8 transformation, we observe a similarly strong temperature dependence of the resulting BC8 microstructure: at high temperatures slightly below the BC8-supercooled liquid boundary, single crystal BC8 is formed. At lower temperatures, more grains are observed, and their dimensions decrease. Figure 3a illustrates the diversity of microstructures at pressure of 1, 800 GPa and temperatures between 3, 500 and 5, 500 K. At temperatures below 3, 500 K, we do not see nucleation of BC8 as the liquid transforms to amorphous carbon characterized by a lack of mobility of carbon atoms. Figure 3b illustrates the growth of BC8 nanocrystal from carbon melt at 1,800 GPa and 4,900K (see also the Movie SM-2 in the Supporting Information). Within the first 8ps, multiple BC8 clusters form and grow, with new clusters appearing. At 20 ps the entire sample is converted to fully solidified nanocrystalline BC8.

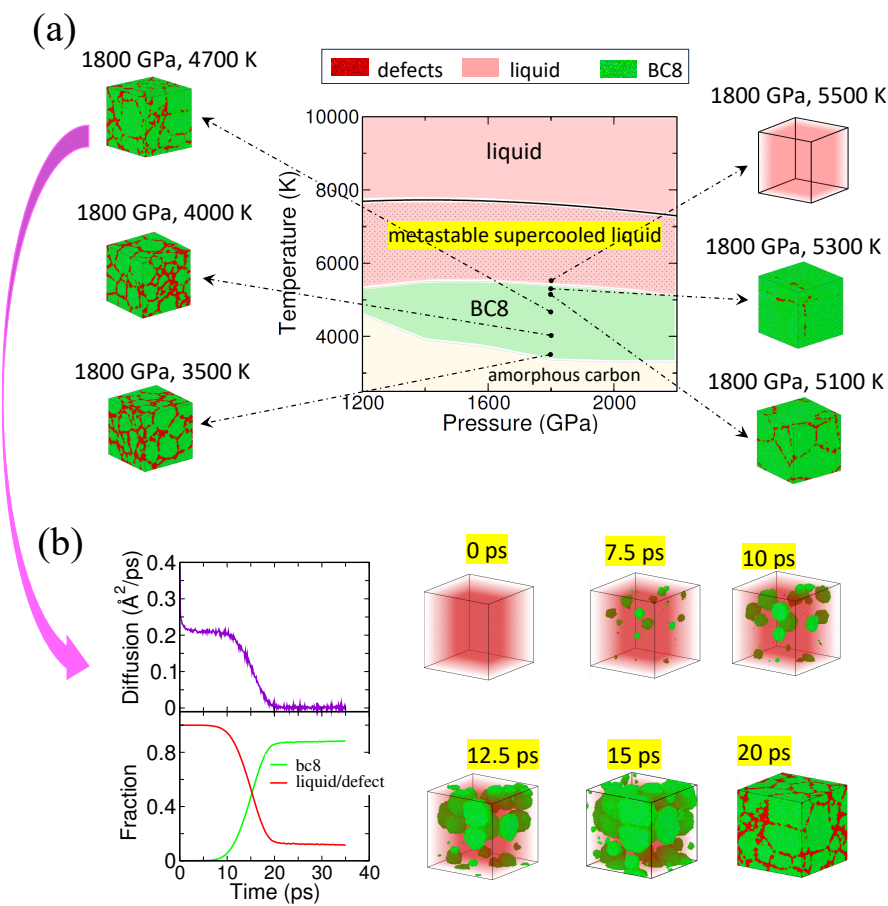

Figure 3: (a) Pressure-temperature map of the final states of supercooled carbon liquid, brought to pressures 1,200 to 2,200 GPa and cooled down to temperatures from 2,800 to 8,000 K, obtained in a series of nanosecond NPT SNAP MD simulations. Carbon stays in metastable liquid state at temperatures below BC8 melting line down to 5,300K. Below this temperature, BC8 nucleates and grows. At even lower temperatures (3,300 − 4,500 K), amorphous carbon is formed as evidenced by the lack of appreciable carbon diffusion. The frames on the left and right show the final states at 1,800 GPa as a function of temperature: metastable liquid at 5,300 K, and BC8 at all other temperatures. The BC8 microstructure displays an increasing number of grains upon reduction of temperature. (b) Time evolution of metastable liquid carbon sample towards nano-crystalline BC8 at 1,800 GPa and 4,700 K. The diffusion coefficient and liquid fraction on the left clearly show the completion of solidification at 20 ps. The frames on the right illustrate the time progression of nucleation and growth of nanocrystalline BC8 in metastable carbon liquid. See also the Movie SM-2 in the Supporting Information.

To rationalize the existence of the maximum temperature for BC8 nucleation, at least 1,000K below the BC8 melting line, we apply classical nucleation theory (CNT)[47–49] with its key parameters (nucleation barrier, critical nucleus size, interfacial, bulk, and liquid free energies, and Zeldovich factor) obtained from atomistic MD simulations (Figure 4). In addition, CNT allows us to estimate the upper temperature bound for an experimentally relevant system size ($\sim 10^{19}$ atoms), which is still many orders of magnitude larger than what can be simulated with MD.

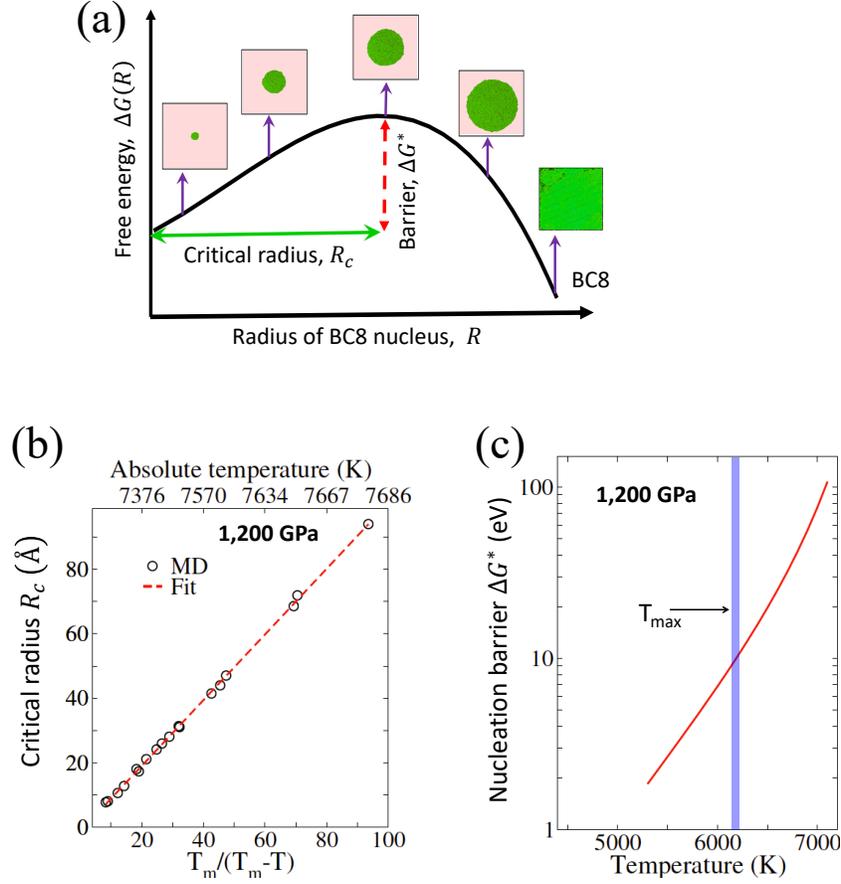

Figure 4: Classical nucleation theory applied to calculate the maximum temperature of BC8 synthesis. (a) Illustration of the concept of critical nucleus and the nucleation barrier ΔG*. (b) Critical nucleus radius $R_c$ as a function of inverse of normalized supercooling $(T_m − T)/T_m$. (c) Nucleation barrier ΔG* as a function of temperature. Blue vertical line displays the maximum temperature $T_{max}$ for BC8 nucleation in a typical NIF target of dimensions 1mm×1mm×100μm at 1,200 GPa.

According to CNT, the nucleation process, involving the appearance and disappearance of small solid clusters in the metastable supercooled liquid, becomes irreversible when the clusters exceed the critical size $R_c$ (Figure 4). The formation free energy, ΔG(R), of a cluster of radius R, is the sum of two contributions: the bulk energy gain to form the solid nucleus from supercooled liquid and the work needed to create the interface between solid and liquid:

$$\Delta G(R) = \frac{4}{3}\pi R^3 \Delta g_v + 4\pi R^2 \sigma_s \quad (1)$$

Here, $\Delta g_V$ is the Gibbs free energy difference between the solid and liquid phases ($\Delta g_V$ and $\sigma_s$ is the interfacial free energy. The critical nucleation radius is then

$$R_c = -2\sigma_s/\Delta g_V, \quad (2)$$

which maximizes ΔG:

$$\Delta G^* = \Delta G(R_c) = \frac{16\pi}{3}\frac{\sigma_s^3}{\Delta g_v^2} \tag{3}$$

and defines the probability of forming a critical nucleus, $\exp(-\Delta G^*/k_B T)$, which then grows irreversibly to form a new phase (Figure 4a). At a given (P,T) point, $\Delta g_V$ is determined using thermodynamic integration[45,46]. The critical nucleus size $R_c$ as a function of temperature is obtained directly from a series of constant enthalpy (NPH) MD simulations of a solid nucleus embedded in the supercooled liquid[50,51] (Figure 4b). The interfacial free energy is calculated using Eq. (2) with the computed $\Delta g_V$ and $R_c$. It was found that the critical nucleus size is a linear function of the inverse of normalized supercooling $(T_m - T)/T_m$, which is the major driving force for the nucleation of a new phase. The corresponding nucleation energy barrier is shown in Figure 4c as a function of temperature. In the limit of $T \to T_m$, the nucleation barrier $G^*$ becomes infinite as $\Delta g_V \to 0$. This explains a significant degree of supercooling (in our case $T_m - T \sim 1,000$ K) required to drive irreversible nucleation of the new phase.

We aim to predict the (P,T) conditions for robust synthesis of BC8 phase in dynamic compression experiments at NIF. For this purpose, we estimate the maximum temperature, $T_{max}$, for BC8 nucleation at a given pressure using the CNT expression for the number of critical nuclei $N^* = N_0/N_{at}^C Z \exp(-\Delta G^*/k T)$[49], where $N_0$ is number of available nucleation sites, $N_{at}^C$ is the average number of atoms in the critical nucleus, and Z is the Zeldovich factor, which accounts for the probability that a nucleus on top of the barrier $\Delta G^*$ proceeds downhill to form the new phase rather than to be dissolved (Figure 4a). Statistics from multiple MD simulations of BC8 nucleation establishes that $N_c > 10$. To determine the upper temperature bound, we estimate $N_0$ as 1/10 of the total number of atoms in a compressed target with dimensions 1 mm × 1 mm × 100 μm, $N_0 \approx 10^{18}$ atoms. As an example, at 1,200 GPa, a value of $N_{at}^C \cong 130$, which gives $T_{max} \cong 6,300$ K (Figure 4c). This is 1,000 K above the maximum at max temperature of BC8 synthesis from carbon melt, meaning that BC8 nucleation should be possible in a potential NIF shot at 1,200 GPa as long as the temperature is within the green region labeled BC8 (Figure 5).

Once the maximum temperature of BC8 nucleation is established for each pressure, we proceed with construction of the final pressure-temperature map of metastable and stable end states of carbon, which emerge upon compression and heating of the NCD carbon precursor (Figure 5). We combine the results of direct MD simulations of NCD including pre-melting and BC8 formation, MD simulations of BC8 nucleation from metastable supercooled carbon liquid, and CNT predictions of BC8 nucleation. The direct MD simulations provide the pre-melting line of NCD as the lowest temperature bound for BC8 synthesis. We predict that without the appreciable volume of metastable carbon liquid it is impossible to initiate BC8 nucleation. Therefore, the NCD sample has to be pre-melted to initiate the transformation. The BC8 nucleation from metastable supercooled liquid provides an estimate of the lowest pressure boundary, 1,200 GPa. The upper temperature boundary of the P − T region for BC8 nucleation is calculated using CNT. The predicted region of BC8 synthesis is larger than that from direct MD simulations. Although we use very large simulation samples (up to 1 billion atoms), their size is still many orders of magnitude smaller than the experimental samples, and the simulation time (up to 1 ns) is still at least an order of magnitude shorter of the time of experiments. Utilizing CNT and MD nucleation simulations from liquid allow us to partially address these time and length scale limitations in a physically relevant way.

In this work, we have uncovered the extreme metastability of diamond, extending substantially into the P − T space of BC8 thermodynamic stability. The extreme metastability stems from the remarkable strength of carbon-carbon bonds, rather than being dependent on MD simulation time. This conclusion is substantiated by employing CNT to extrapolate to experimental time and length scales. This important observation explains the extreme difficulty of experimental synthesis of BC8, as the predicted (P − T ) region of BC8 synthesis is highly localized, requiring pressures above 1,200 GPa and temperatures between 3,700 and 6,300K (green area of Figure 5). The challenge is then to reach this BC8 synthesis region in experiments capable of documenting the atomic-scale structure.

In contrast to ramp compression, the conservation of mass, momentum and energy in shock compression defines a unique compression pathway, which is referred to as the shock Hugoniot[52]. Previous experiments suggest that the single-shock Hugoniot reaches the diamond melting line at 700 GPa and exits into liquid around 1, 000 GPa[29]. Therefore, the BC8 synthesis region cannot be reached in single-shock compression of a diamond sample. Double-shock compressions, i.e. loading a sample with an initial shock before launching a faster, stronger second shock in the material that is already compressed and heated by the initial shock, allows one to reach lower temperature states compared to those accessible with a single shock (Figure 5).

We design double-shock compression pathways to reach the region in P − T space where we predict the synthesis of BC8 (green area in Fig. 5). The Hugoniots, the loci of final thermodynamic states of a material behind a series of individual shocks (blue and red diamonds in Figure 5), are calculated in a series of NPT MD simulations: at a given pressure P, temperature T is varied to satisfy the Hugoniot energy conservation equation:

$$E(T,P) - E(T_0, P_o) + \frac{1}{2}(P + P_0)(V_0 - V(T,P)) = 0 \qquad (4)$$

where E(T, P ) is the specific internal energy, and V (T, P ) = 1/ρ(T, P ) is the reduced volume, and ρ is the density of the material. The first shock starts at $(T_0, P_0)$ at 300K and 0 GPa. The starting $(T_0, P_0)$ point of 2nd Hugoniot is iteratively chosen on the 1st Hugoniot to reach the final (T , P ) state with minimum pressure around 1, 200 GPa within the green area in Figure 5. The starting material is an NCD sample, which gradually evolves along the 2nd Hugoniot to a pre-melted, coarsened state at 1, 200 GPa. As discussed above, although we cannot directly simulate the BC8 appearance at this pressure due to time and length scale limitations of MD, CNT predicts the nucleation of BC8, which we aim to observe in our NIF experiments. The final pressure 1, 200 GPa was chosen to satisfy an experimental requirement at NIF. To reliably observe X-ray diffraction signal of BC8, the final pressure should as low as possible to reduce background radiation from the ablation plasma. The double-shock pathway shown in Figure 5 follows a 1st shock to (350 GPa, 2, 950 K), followed by a 2nd shock to (1, 200 GPa, 6, 019 K).

In a recent publication[53], Shi et al. proposed double shock compression pathway to BC8 phase of carbon based on the so-called Multi-Scale Simulation Technique (MSST) simulations. MSST is an extended Lagrangian MD method developed to reach the final Hugoniot state, by compressing material along a straight line in pressure-volume space, known as the Rayleigh line, through fictitious strain-rate dynamics[54]. Using explicit, piston driven MD shock simulations, we attempted to corroborate their results but were unsuccessful. We predict that the final state of the double shock compression to 1, 713 GPa is indeed a liquid, but its temperature of 5,756K, slightly above the diamond melting line, is appreciably higher than the temperature of the intermediate liquid compression state ( 5,136K), from which Shi at al. observed the BC8 nucleation. As shown in this work (Figure 2a), under such conditions (1, 713 GPa and 5, 756 K), BC8 cannot be nucleated at nanosecond time scale of MD simulations. Given we were unable to confirm the results presented by Shi et al. with direct shock simulations (deemed superior to MSST), we believe their findings are likely an artifact of MSST. Details of our direct MD shock simulations are provided in the Supporting Information.

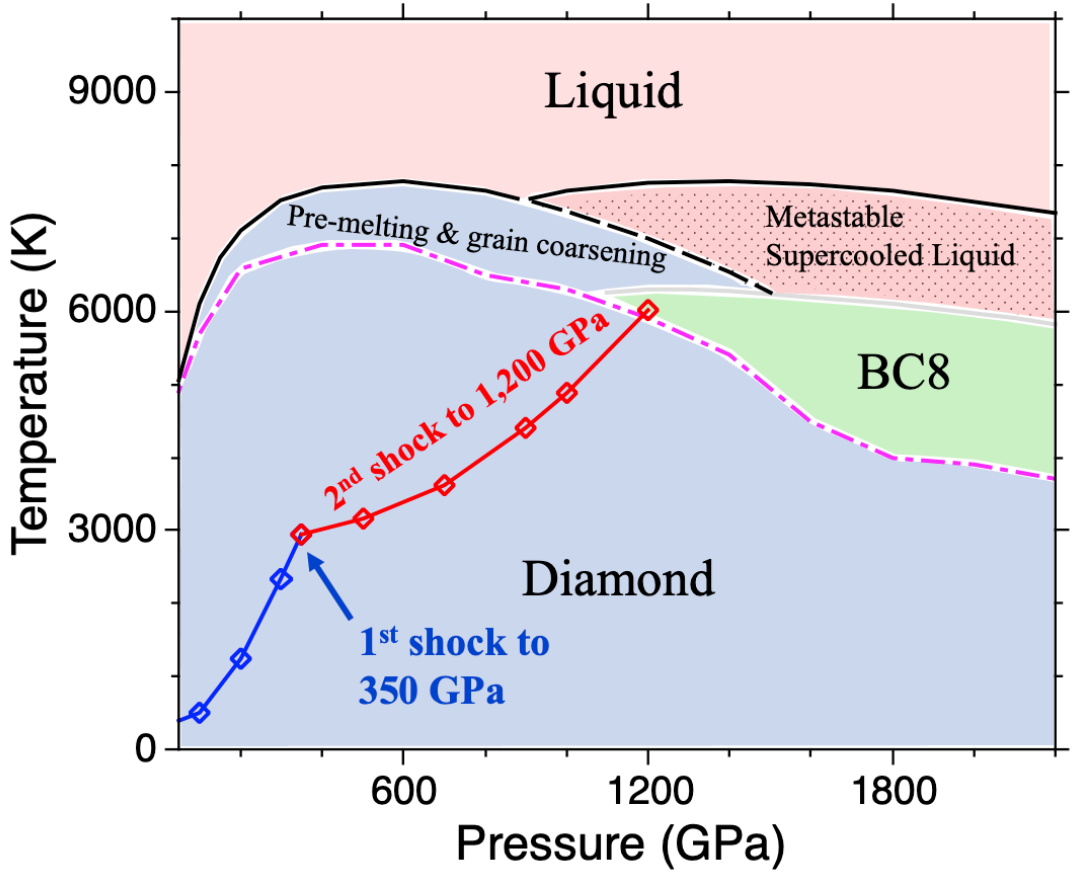

Figure 5: Final pressure-temperature map of metastable and stable end states of carbon, which appear upon compression and heating of NCD carbon precursor constructed by combining direct MD simulations of NCD including pre-melting and BC8 formation (Figure 2), MD simulations of BC8 nucleation from metastable supercooled carbon liquid ( 3), and CNT prediction of BC8 nucleation (Figure 4). The double shock compression pathway to reach minimum pressure for BC8 synthesis, 1,200 GPa. First shock (blue line) brings NCD to 350 GPa and 3,000 K, followed by the second shock (red line) to the final point at 1,200 GPa and 6,000 K.

The prediction of the BC8 synthesis by our quantum accurate MD simulations laid the foundation for a successful proposal to NIF Discovery Science program at Lawrence Liver- more National Laboratory, which was designed to test our predictions, originally reported in Ref.[43]. We are currently fielding experiments at NIF based on the double-shock design suggested by SNAP MD simulations[55].

In summary, by utilizing the transformative capability of quantum-accurate molecular dynamics simulations enabled by the SNAP machine learning interatomic potential, we investigated the materials transformations of nanocrystalline diamond within a wide range of pressures and temperatures. We observed its extreme metastability extending well be- yond the diamond/BC8 phase boundary. The elusive, but long sought BC8 phase of carbon nucleates and grows upon pre-melting of NCD into metastable supercooled carbon liquid. While multiple studies predicted that BC8 is the thermodynamically stable form of carbon between 1 and 3 TPa, our simulations reveal that the $P-T$ domain of the carbon phase diagram, where BC8 could be experimentally synthesized, is much smaller. We, therefore, demonstrate that experiments must achieve the $P-T$ conditions within this specific synthesizability domain to observe the BC8 post-diamond phase of carbon. This finding aligns with the persistence of FC8 (diamond) up to 2 TPa observed in recent ramp-compression experiments of Lazicki et al.[33]. Using SNAP MD, we constructed the double-shock com- pression pathway, which is currently being explored in theory-inspired experiments at the National Ignition Facility aimed to synthesize the BC8 phase of carbon.

## Supporting Information

The supporting information is available free of charge at https://pubs.acs.org/doi/

- Additional details including methods, NCD sample preparation, supporting Figures S1-S4 (**pdf**).

- Movie SM-1: Transformation of NCD sample through full melting and solidification to BC8 (**mov**)

- Movie SM-2: nucleation and growth of nanocrystalline BC8 in metastable carbon liquid (**mov**)

## Corresponding Author

*Ivan I. Oleynik - Department of Physics, University of South Florida, Tampa, FL 33620 Email: oleynik@usf.edu*Ivan I. Oleynik - Department of Physics, University of South Florida, Tampa, FL 33620 Email: oleynik@usf.edu

## Acknowledgement

The work at USF is supported by DOE/NNSA (Award Nos. DE-NA-0003910 and DE-NA-0004089) and DOE/FES (Award Nos. DE-SC0023508 and DE-SC0024640). Sandia National Laboratories is a multi-mission laboratory managed and operated by National Technology and Engineering Solutions of Sandia, LLC, a wholly owned subsidiary of Honeywell International, Inc., for the U.S. Department of Energy's (DOE) National Nuclear Security Administration under Contract No. DE-NA0003525. The part of this work was performed by Lawrence Livermore National Laboratory under Contract No. DE-AC52-07NA27344. The computations were performed using leadership-class HPC systems: OLCF Summit and Frontier at Oak Ridge National Laboratory (ALCC and INCITE Awards Nos. MAT198 and MAT261), ALCF Polaris at Argonne National Laboratory (ALCC Award "FusAblator") and TACC Frontera at University of Texas at Austin (LRAC Award No. DMR21006).

**Notes**

The authors declare no competing financial interest.